\documentclass[]{raa}            

\usepackage{graphicx,times}             
\usepackage{natbib}
\usepackage{amssymb,amsmath}
\bibpunct{(}{)}{;}{a}{}{,}

\usepackage[a4paper=true,pagebackref=true]{hyperref}
\hypersetup{colorlinks = true, linkcolor = green, anchorcolor = red, citecolor = blue, filecolor = red, pagecolor = red, urlcolor = red}

\begin{document}

   \title{Variable binaries and variables in binaries in the Binary Star Database BDB
}

   \volnopage{Vol.0 (20xx) No.0, 000--000}      
   \setcounter{page}{1}          

   \author{D. Kovaleva
      \inst{}
   \and O. Malkov
      \inst{}
   \and P. Kaygorodov
      \inst{}
   }

   \institute{Institute of Astronomy, Russian Academy of Sciences,
             Moscow 119017, Russia; {\it dana@inasan.ru}\\
\vs\no
   {\small Received~~20xx month day; accepted~~20xx~~month day}}

\abstract{ The BDB, Binary star DataBase
\url{http://bdb.inasan.ru}
combines data of the catalogues of binary and multiple stars of all observational types. There is a number of ways for variable stars to form or to be a part of binary or multiple systems. We describe how such stars are represented in the database.}


   \authorrunning{D. Kovaleva, O. Malkov, \& P. Kaygorodov }            
   \titlerunning{Variable binaries and variables in binaries in the BDB}  

   \maketitle

%
%
\section{Introduction}           
\label{sect:intro}

Binary stars are very numerous, and binarity rate increases from $\sim 20\%$ for M-dwarfs up to $\approx 60-70\%$ for A- and earlier spectral type stars  (\citealt{2013ARA&A..51..269D}). Besides, in many cases (probably as often as $1/4$, according to \citealt{2014AJ....147...87T}) at thorough investigation, binaries prove to be multiples. Information on properties of population of binary and multiple stars has significant influence on studies of stellar formation, and of Galaxy evolution (\citealt{2015ASPC..496...37B, 2009A&A...493..979K, 2009MNRAS.393..663W, 2001MNRAS.321..149M, 1991A&A...247...87P}). Binaries exhibit many astrophysical phenomena of extreme interest, such as novae, type I supernovae, Wolf-Rayet stars, X-ray and symbiotic binaries; later stages of evolution of close binaries involve formation of compact objects, such as neutron stars and black holes, including recently discovered events of their merging (\citealt{2017PhRvL.119p1101A, 2017ApJ...850L...1L, 2016PhRvL.116f1102A, 1996MNRAS.283L..37V, 1994MNRAS.268..871T, 1993MNRAS.260..675T}). Moreover, gravitational interaction between components in binary and multiple stars allows us to directly determine physical characteristics of the system and its components, including dynamical mass. This provides calibration of a number of relations between observed and physical parameters applied to single stars (\citealt{2015ApJ...807....3K, 2015AJ....149..151H, 2007MNRAS.382.1073M, 2006DDA....37.1203B, 1997A&A...320...79M}), and tests of models of stellar evolution (see, for instance, \citealt{2017A&A...608A..62H, 2014NewAR..60....1S, 2012A&A...540A..64V, 2002ARep...46..233K}).

Binaries are observed by various
methods resulting in separate datasets named {\it observational types} (\citealt{2017ASPC..510..360M})
(e.g. eclipsing, visual, spectroscopic, etc.). Observational types are determined by observational techniques used to register stellar duplicity, but overlap in the spaces of physical and observational parameters (\citealt{2015BaltA..24..446K}). Usually catalogues and databases address one or few selected observational types. Combining their information may be difficult because of a number of problems, including insufficient or indefinite identifiers, and variety of non-homogeneous data, which can characterize either a component, a pair, or a system as a whole.

The purpose of the Binary star DataBase, BDB \url{http://bdb.inasan.ru} was to combine data of all catalogues and databases of binary and multiple stars, to allow user to obtain complete information about certain star, or about a set of binaries of pre-defined characteristics. This required accurate cross-identification and establishment of proper relations between data and objects in binaries and multiples (\citealt{2015A&C....11..119K, 2012BaltA..21..309K}). The task is solved with an index catalogue of binary and multiple stars ILB (Identification List of Binaries, \citealt{2016BaltA..25...49M}) implementing a specially developed designation scheme BSDB (Binary Star DataBase) with three categories of objects, System, Pair and Component (\citealt{2015BaltA..24..185K}).

The term ``variable stars'' refers to a numerous and diverse sets of stars based on the only observational phenomenon, photometric variability, and relating to a variety of astrophysical processes. If a binary star is observed as a variable, or a variable star is observed in a binary or multiple system, this often multiplies available astronomical information. BDB includes a number of catalogues having data on variable stars of different types, or completely aimed to variable binaries. In this paper, we discuss integration of these catalogues in BDB and related issues.

There are three types of relation between variables and binary (multiple) stars: (i) one or more components of a binary may vary its brightness being a pulsator, or a spotted star, or an erupting star; (ii) the brightness of a pair may vary because of its orbital movement (e.g. eclipsing binaries); (iii) the binarity may cause physical variability (e.g. cataclysmic binaries). Though these types are not mutually excluding (see detailed discussion by \citealt{2005Ap&SS.296..145S}), it is convenient to start discussing them separately. The primary catalogue for all variable stars in BDB is the General Catalogue of Variable Stars (GCVS, \citealt{2017ARep...61...80S}), however there are also catalogues in BDB addressing different observational types separately.

Section 2 discusses binary stars with variable components, Section 3 deals mainly with catalogues of eclipsing binaries, and Section 4 contains information on catalogues of intrinsic binary variables. Section 5 discusses some issues of overlapping observational types and classes of variability, and Section 6 summarizes the results.

\section{Variable stars in binary and multiple systems in BDB}

In this section, we address to cases when one or more components of a binary or multiple star are intrinsic variables. The catalogues of such systems containing few hundreds of stars were compiled by \citealt{1962JO.....45..117B}, \citealt{1963PZ.....14..357P} and \citealt{1989A&AS...77..497B}. Regretfully neither of these catalogues is available in electronic form, so they can be used for a referral but are not integrated in the BDB so far. The data on 171 classical Cepheides in binary and multiple system from the ``Binary stars among Galactic classical Cepheids'' database (\citealt{2003IBVS.5394....1S}) has been integrated in BDB, and a special flag (in Observational Type field) was assigned to all stars from that dataset. For the other types of intrinsic variable stars in binaries, no specially devoted catalogues are available except the ``Candidates for Binaries with an RR Lyrae Component'' (\citealt{2016MNRAS.459.4360L}), which will not be integrated in BDB until the candidates are confirmed.

A catalogue of UV Ceti type flare stars in binary systems was constructed by \citealt{2014AcA....64..359T}. It contains information on 138 such binaries, and data on orbital elements are available for 31 of them.

Some major catalogues contain information on variable components of binaries in text files of comments. In BDB, we work on reading data on properties of binaries from text files. The success strongly depends on whether authors use formal phrases to describe structure of a system. We obtain part of information of stellar types from text comments (Notes) to the Washington Visual Double Star Catalogue (WDS, \citealt{2016yCat....102026M}). In particular, 73 binary and multiple systems from this catalogue are marked to contain pulsators like Cepheide or RR Lyr, though one should note that not necessarily all of these systems are physical (due to principles of this catalogue construction). On the contrary, information on intrinsic variability of the components of systems listed in the Multiple Star Catalogue (MSC, \citealt{2018ApJS..235....6T}) and in The ninth catalogue of spectroscopic binary orbits (SB9, \citealt{2004A&A...424..727P}) definitely refers to physical binaries. GCVS contains valuable information on variables having components in comments, though this information relates mainly to variable stars, and we do not use it in BDB so far. If a binary system is found in GCVS, we mark this systems as one containing an intrinsic (pulsating, eruptive, spotted, etc.) variable, based on the GCVS variability type. However, the BDB information about different types of intrinsic variability is certainly very incomplete. We introduce ``intrinsic variable inside'' type only if a star has a GCVS identifier in some of the catalogues of binaries, and GCVS provides information concerning relevant variability type.

Figures~1 -- 4 below show the screenshots of (some part of) BDB representation for a remarkable multiple system BSDB J194448.73+291552.8:s including a classical Cepheide, SU Cyg. One of the components (BSDB J194448.73+291552.8:c1B) of a spectroscopic binary with an orbital period of $\sim 549$ days, SBC9 1172 = BSDB J194448.73+291552.8:p1A-1B, is the Cepheide, while another component itself is a close spectroscopic binary, SBC9 2142 = BSDB J194448.73+291552.8:p1AA-p1Ab, having orbital period of $\sim 4.7$ days. The wider pair can be resolved interferometrically. Three more distant visual components of yet unknown physical connection are situated at angular separations of $\sim 25, \sim 71, \sim 75$ arcsec, respectively. Combined with parallax $\varpi=1.52\pm0.27$ mas  (Gaia DR1 TGAS, \citealt{2016A&A...595A...4L}), or $\varpi=1.1695 \pm 0.0516$ mas (Gaia DR2, \citealt{2018arXiv180409366L}), this does not favor expectations of orbital motion. On the other hand, based on Gaia DR2 parallaxes for these remote components ($1.2351 \pm 0.0278, 1.9440 \pm 0.0246, 1.1280 \pm 0.0345$ mas), at least two of them are situated rather close to SU Cyg in space, and at least one of them, the one with $\varpi = 1.1280 \pm 0.0345$ mas, has similar proper motion. Anyway, the system is quite interesting and promising, thanks to admirable combination of multiplicity and variability of one companion.

  \begin{figure}
   \centering
   \includegraphics[width=\textwidth, angle=0]{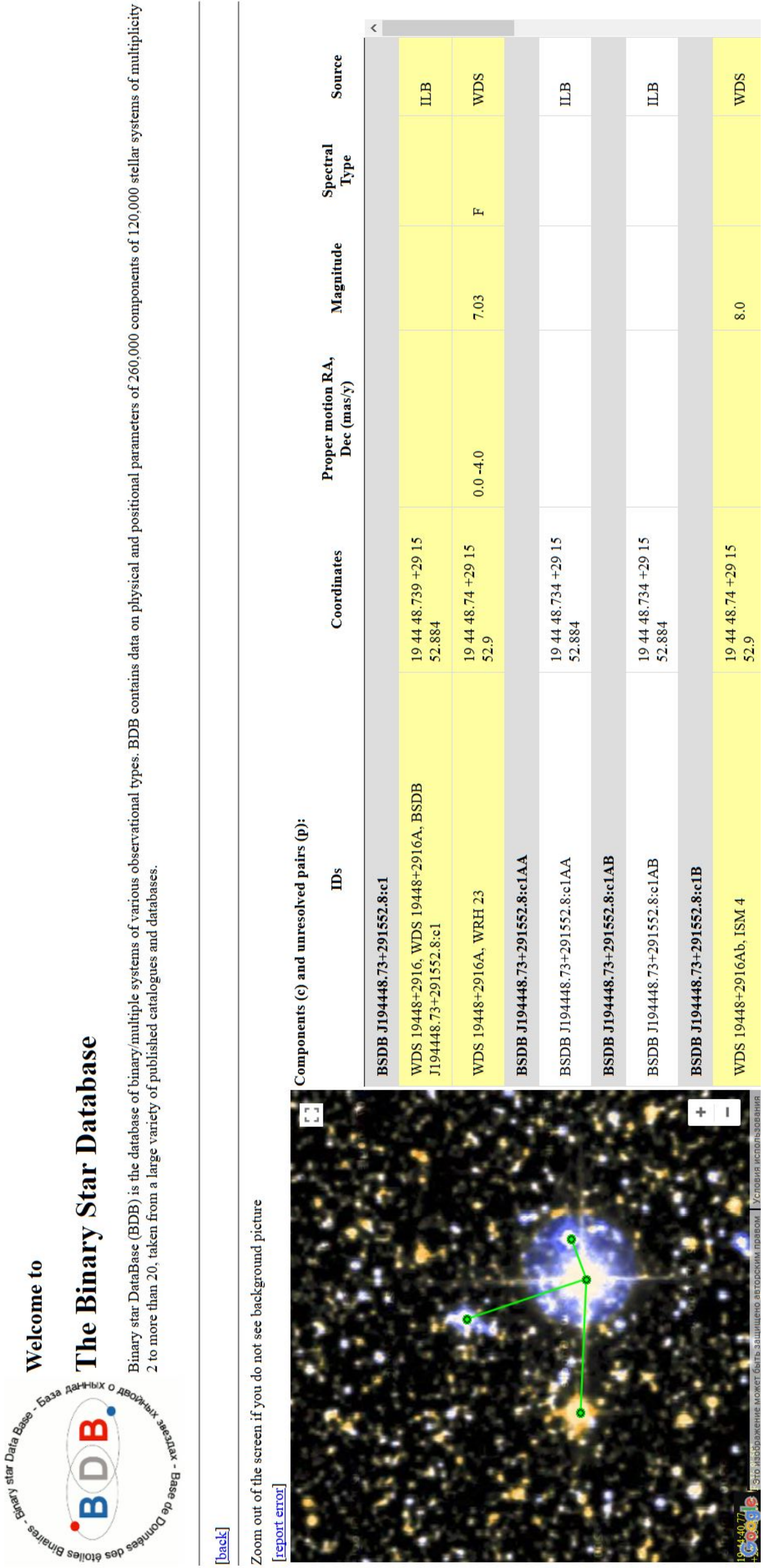}
   \caption{Result of request on ``SU Cyg'' in BDB. Visualization of the multiple system including Cepheide, and part of data on components. Yellow background marks lines referring to requested identifier, other (white) lines represent information for other objects in the same system. }
   \label{Fig1}
   \end{figure}

 \begin{figure}
   \centering
   \includegraphics[width=\textwidth, angle=0]{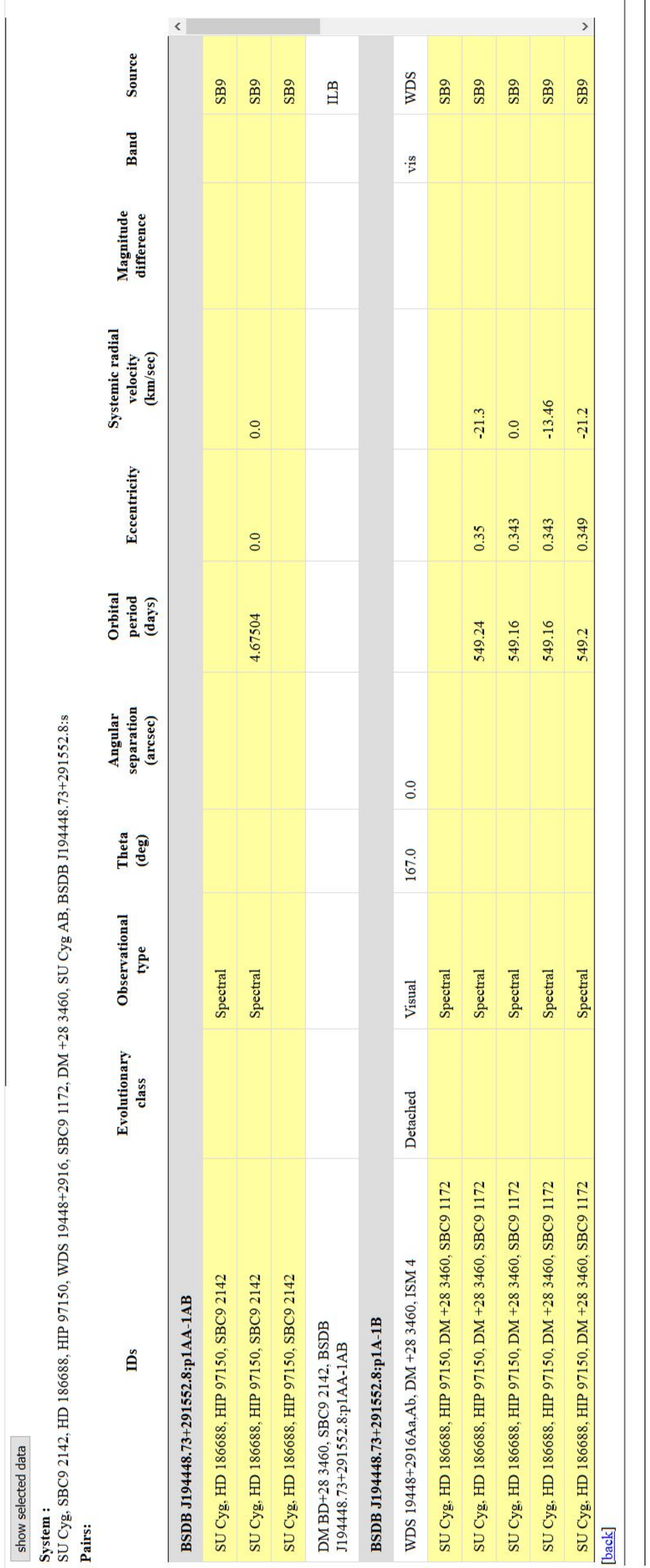}
   \caption{Result of request on ``SU Cyg'' in BDB. List of identifiers and part of data on pairs. The ``Show selected data'' button in the upper part of figure (just below the visualization figure on the real screen) leads to lines of original data from the catalogues for objects on yellow background.}
   \label{Fig2}
   \end{figure}

 \begin{figure}
   \centering
   \includegraphics[width=\textwidth, angle=0]{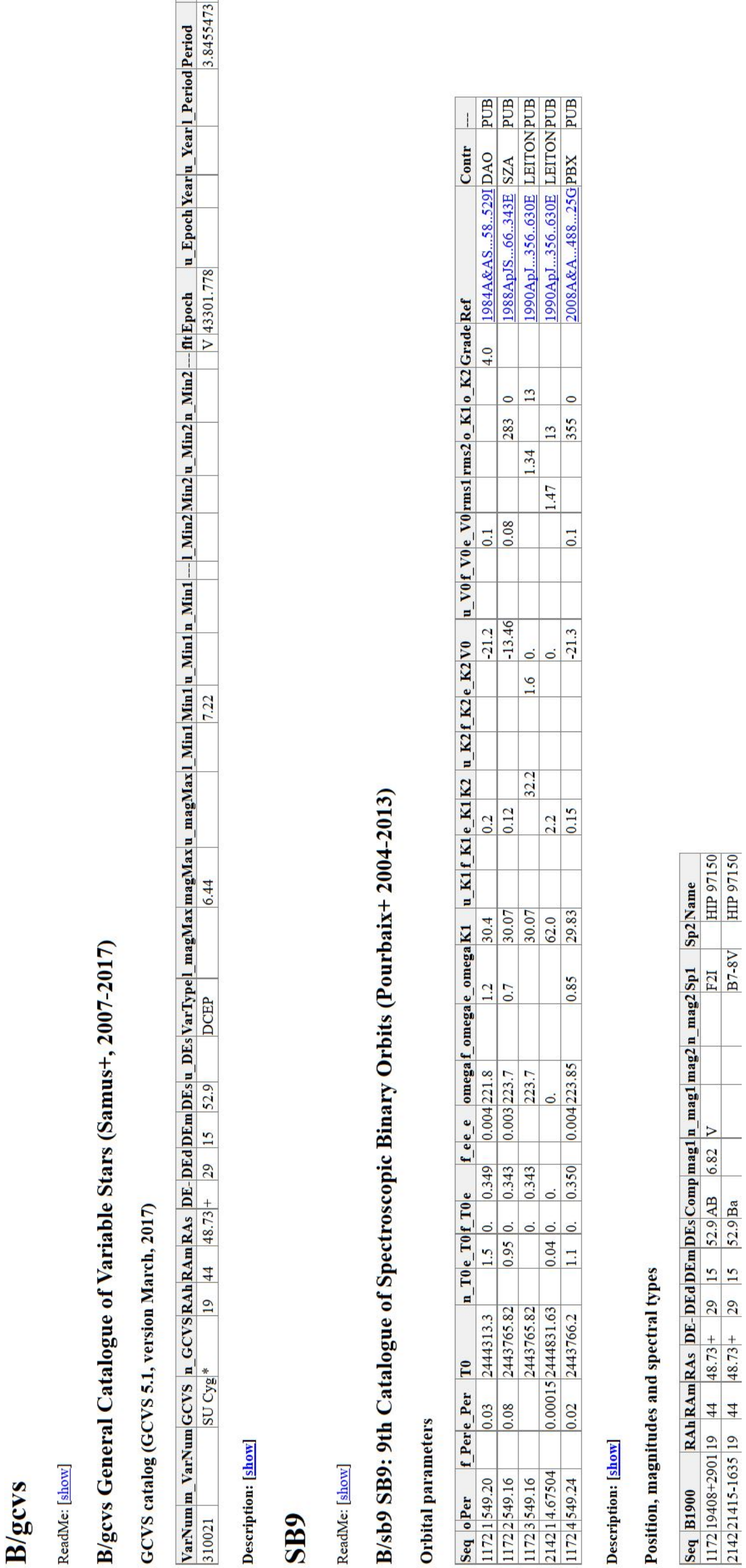}
   \caption{One part of result of request on ``SU Cyg'' in BDB, second level (after the ``Show selected data'' button is pressed). Original data from catalogues.}
   \label{Fig3}
   \end{figure}

 \begin{figure}
   \centering
   \includegraphics[width=\textwidth, angle=0]{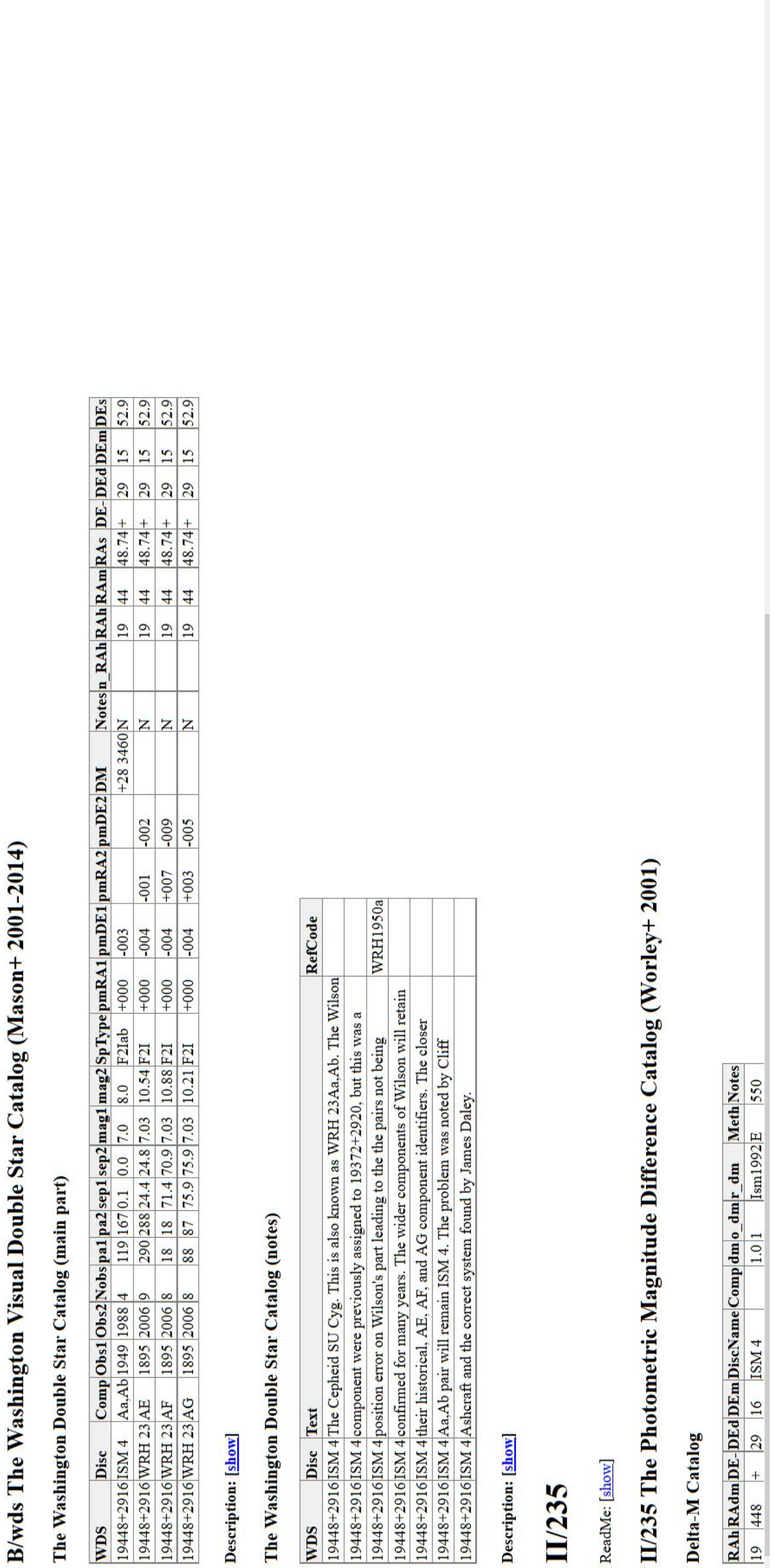}
   \caption{Other part of result of request on ``SU Cyg'' in BDB, second level (after the ``Show selected data'' button is pressed). Original data from catalogues (not all catalogues represented).}
   \label{Fig4}
   \end{figure}

\section{Eclipsing binaries catalogues}

Eclipsing binaries form the most populated observational type of binaries, hundreds of thousands of such stars are already discovered in our Galaxy and in Magellanic Clouds. The increase of number of known eclipsing binaries during recent decades thanks to projects searching for brightness variability (ASAS, \citealt{2006MNRAS.368.1311P}) due, in particular, to gravitational microlensing or exoplanet transit (OGLE, \citealt{2016AcA....66..405S}; Kepler, \citealt{2016AJ....151...68K}; et al.).
Numerous eclipsing binaries discovered in the course of such projects cannot, as a rule, be individually investigated and classified. In the meantime, the recent version of GCVS (\citealt{2017ARep...61...80S}) includes more than 10000 eclipsing variables having sufficiently complete data.

Eclipsing binaries can be very different both morphologically and physically. There is a number of catalogues devoted to some subclasses of these stars in BDB. In respect with other observational types of stars, catalogues devoted to reliably classified eclipsing binaries contain more physical information on the systems and components. In particular, large volume of physical data was obtained by M.A.Svechnikov and his colleagues in their investigation of lightcurves of eclipsing variable stars (\citealt{2004yCat.5118....0P}, \citealt{2004yCat.5115....0S}, \citealt{2005A&A...437..375D}, \citealt{1984BICDS..26...99S}, \citealt{1996OAP.....9...20B}, \citealt{2011yCatp044002301P}, \citealt{2004yCat.5124....0S}).

The Catalogue of Eclipsing Variables (CEV, \citealt{2013AN....334..860A}, \citealt{2006A&A...446..785M}) contains the most complete data for eclipsing variables of known morphological types. It combines data from GCVS tables and text remarks with data of other bibliographic sources and observations, using author's original scheme of lightcurve processing to determine physical type of a binary and its characteristics.

The gap between a number of discovered eclipsing binaries and possibilities of individual investigation of every star to find out its physical characteristics makes automatization of lightcurves classification extremely actual (see, for instance, \citealt{2018arXiv180109406M}, \citealt{2016MNRAS.456.2260A}, \citealt{2014MNRAS.444.1982A}, \citealt{2008ASPC..394..381M}). It is expected that as one of results of Gaia mission, several millions of eclipsing binaries will be discovered  (\citealt{2017A&A...603A.117S}, \citealt{2017A&A...602A.110K}).

So far the extensive lists of recently discovered eclipsing binaries were not included into BDB, due both to their underclassification and lack of chances that they may intersect with any other catalogues. However, recent progress in automated classification of lightcurves, systematic investigations of eclipsing binaries discovered in large surveys with different techniques (\citealt{2016AcA....66..405S}, \citealt{2016PASP..128g4201K}, \citealt{2009MNRAS.395..593P}), objects observed in more than one survey (\citealt{2014ApJ...790..157D}, \citealt{2014MNRAS.443..432M}, \citealt{2013A&A...553A..18S}), many objects of interest among these eclipsing binaries (see, e.g., eclipsing binaries with components-Cepheides, \citealt{2016pas..conf...31P},  \citealt{2015AcA....65..341U};  spectroscopically resolved eclipsing binaries \citealt{2004A&A...426..577M}, \citealt{2016PASP..128g4201K}) make us start preparation for integration of extensive lists of eclipsing binary stars, discovered in surveys, into BDB index catalogue, ILB.

\section{Catalogues of intrinsic binary variable stars}

In GCVS, there is a number of classes of physically variable stars with variability caused by their binarity. Let us name them intrinsic binary variable stars.

Currently about 2000 cataclysmic variable stars are known. In addition to GCVS, two main (overlapping) catalogues for this observational type are those by \citealt{2011yCat....102018R} (was regularly updated up to 2016), and by \citealt{2006yCat.5123....0D}. The data for these binaries are highly non-uniform, but usually include orbital characteristics (period, inclination limitations), morphological type of a variable, and relative characteristics of the components.

Some binaries demonstrate variability out of optical range. In BDB, there are data on about 300 X-ray binaries (partly included in the already mentioned catalogue by \citealt{2011yCat....102018R}). Other data sources for these objects are Catalog of Galactic Low-Mass X-Ray Binaries (LMXB, \citealt{2007A&A...469..807L}), and Catalog of Galactic High-Mass X-Ray Binaries (HMXB, \citealt{2006A&A...455.1165L}). Besides, there are more than 200 radiopulsars discovered in binary systems. These objects were identified with optical sources for their integration into BDB index-catalogue ILB and, therefore, into the database (\citealt{2015BaltA..24..395M}).

We do not yet include data on the classes of binaries discovered on the stage, when they actually terminate their binary, like SN Ia and gravitational-wave binaries.

There is a group of variable binaries of brightness varying with stellar or orbital rotation. Simultaneously, this variability critically depends also on their binarity: these are ellipsoidal, reflecting, spotted binaries (\citealt{2005Ap&SS.296..145S}). GCVS is the main source of data for these stars in BDB.


\section{Conclusions}
\label{sect:conclusion}

Variable stars of different physical and observational nature are represented in BDB by the data of a number of catalogues. As part of binary or multiple systems, intrinsic variable stars can provide more interesting data than single ones. Intrinsic binary variable stars are about the most popular astrophysical laboratories. Eclipsing binary stars form the most populated and rapidly increasing class of variable binaries. Only smaller, well-studied part of them from the GCVS catalogue is presented in BDB so far. These stars allow determination of astrophysical data if their physical nature is properly classified. Automated classification of lightcurves of eclipsing binaries is an important challenge of present and nearest future.

While this paper was almost completed, the 2nd Data Release of Gaia mission (\citealt{2016A&A...595A...1G}) has just been published. Though, contrary to earlier plans, this data release does not include catalogues of non-single stars, it certainly will publish an amount of new data on variable stars. We work on BDB and its index-catalogue, ILB, to implement data of present and coming large catalogues for the benefit of investigations of binary stars, including variable binaries and variables in binaries.

\begin{acknowledgements}
This work was partly supported by the Russian Foundation of Basic Researches, projects 16-07-1162 and 18-02-00890. The support by Fundamental Research Support Program 28 by the Russian Acad. Sci. is acknowledged. This research has made use of the VizieR catalogue access tool, CDS, Strasbourg, France. This work has made use of data from the European Space Agency (ESA) mission {\it Gaia} (\url{https://www.cosmos.esa.int/gaia}), processed by the {\it Gaia} Data Processing and Analysis Consortium (DPAC, \url{https://www.cosmos.esa.int/web/gaia/dpac/consortium}). Funding for the DPAC has been provided by national institutions, in particular the institutions
participating in the {\it Gaia} Multilateral Agreement.
\end{acknowledgements}




\label{lastpage}

\end{document}